\documentclass[10pt]{article}
\usepackage{bbm}
\usepackage[final]{graphics}
\usepackage{amsmath}
\usepackage{amsfonts,amsbsy}
\usepackage{amssymb}

\def\empile#1\over#2{\mathrel{\mathop{\kern 0pt#1}\limits_{#2}}}
\def\bs{\boldsymbol}

\newcommand{\slv}{\raise.15ex\hbox{$/$}\kern-.53em\hbox{$v$}}
\newcommand{\slF}{\raise.15ex\hbox{$/$}\kern-.53em\hbox{$F$}}
\newcommand{\slL}{\raise.15ex\hbox{$/$}\kern-.53em\hbox{$L$}}
\newcommand{\slP}{\raise.15ex\hbox{$/$}\kern-.53em\hbox{$P$}}
\newcommand{\slp}{\raise.15ex\hbox{$/$}\kern-.53em\hbox{$p$}}
\newcommand{\slq}{\raise.15ex\hbox{$/$}\kern-.53em\hbox{$q$}}
\newcommand{\slR}{\raise.15ex\hbox{$/$}\kern-.53em\hbox{$R$}}
\newcommand{\slQ}{\raise.15ex\hbox{$/$}\kern-.53em\hbox{$Q$}}
\newcommand{\slK}{\raise.15ex\hbox{$/$}\kern-.53em\hbox{$K$}}
\newcommand{\slk}{\raise.15ex\hbox{$/$}\kern-.53em\hbox{$k$}}
\newcommand{\slD}{\raise.15ex\hbox{$/$}\kern-.53em\hbox{$D$}}
\newcommand{\slC}{\raise.15ex\hbox{$/$}\kern-.53em\hbox{$C$}}
\newcommand{\slA}{\raise.15ex\hbox{$/$}\kern-.53em\hbox{$A$}}
\newcommand{\slSigma}{\raise.15ex\hbox{$/$}\kern-.53em\hbox{$\Sigma$}}
\newcommand{\slpartial}{\raise.15ex\hbox{$/$}\kern-.53em\hbox{$\partial$}}
\newcommand{\slcalP}{\raise.15ex\hbox{$/$}\kern-.63em\hbox{$\cal P$}}

\def\q{{\boldsymbol q}}
\def\l{{\boldsymbol l}}
\def\k{{\boldsymbol k}}

\def\x{{\boldsymbol x}}

\def\b{{\boldsymbol b}}

\catcode`\@=11

\newcount\@tempcntc
\def\@citex[#1]#2{\if@filesw\immediate\write\@auxout{\string\citation{#2}}\fi
  \@tempcnta\z@\@tempcntb\m@ne\def\@citea{}\@cite{%
        \@for\@citeb:=#2\do%
    {\@ifundefined{b@\@citeb}%
        {\@citeo\@tempcntb\m@ne\@citea%
                \def\@citea{,\penalty\@m\ }{\bf ?}\@warning%
                {Citation `\@citeb' on page \thepage \space undefined}}%
        {\setbox\z@\hbox{\global\@tempcntc0\csname b@\@citeb\endcsname\relax}
     \ifnum\@tempcntc=\z@ \@citeo\@tempcntb\m@ne%
       \@citea\def\@citea{,\penalty\@m}%
       \hbox{\csname b@\@citeb\endcsname}%
     \else%
      \advance\@tempcntb\@ne%
      \ifnum\@tempcntb=\@tempcntc%
      \else\advance\@tempcntb\m@ne\@citeo%
      \@tempcnta\@tempcntc\@tempcntb\@tempcntc\fi\fi}}\@citeo}{#1}}%

\def\@citeo{\ifnum\@tempcnta>\@tempcntb\else\@citea
  \def\@citea{,\penalty\@m}%
  \ifnum\@tempcnta=\@tempcntb\the\@tempcnta\else
   {\advance\@tempcnta\@ne\ifnum\@tempcnta=\@tempcntb \else
\def\@citea{--}\fi
    \advance\@tempcnta\m@ne\the\@tempcnta\@citea\the\@tempcntb}\fi\fi}

\catcode`\@=12

\begin{document}

\title{\bf Distribution of multiple scatterings in\\ proton-nucleus collisions at high energy}
\author{Nicolas Borghini$^{(1)}$, Fran\c cois Gelis$^{(2)}$}
\maketitle
\begin{center}
\begin{enumerate}
\item
Theory Group, Physics Department\\
CERN\\
CH-1211 Geneva 23, Switzerland
\item
Service de Physique Th\'eorique (URA 2306 du CNRS)\\
CEA/DSM/Saclay, B\^at. 774\\
91191, Gif-sur-Yvette Cedex, France
\end{enumerate}
\end{center}

\begin{abstract}
We consider proton-nucleus collisions at high energy in the Color
Glass Condensate framework, and extract from the gluon production
cross-section the probabilities of having a definite number of
multiple scatterings in the nucleus. Various properties of the
distribution of the number of multiple scatterings are studied, and we
conclude that events in which the momentum of a hard jet is
compensated by many much softer particles on the opposite side are
very unlikely except for extreme values of the saturation momentum.
In the same framework, we also investigate the possibility to
estimate the impact parameter of a proton-nucleus collision, from the
measure of the multiplicity of its final state.
\end{abstract}
\vskip 5mm
\begin{flushright}
Preprint CERN-PH-TH/2006-126, SPhT-T06/075
\end{flushright}

\section{Introduction}
Hadronic collisions at high energy involve the interaction of partons
that carry a very small fraction $x$ of the longitudinal momentum of
the incoming projectile. Since the occupation number for such states
in the nucleon wave function can become quite large, one expects that
the physics of parton saturation \cite{GriboLR1,MuellQ1,BlaizM1} plays
an important role in such studies. This saturation generally has the
effect of reducing the number of produced particles compared to what
one would have predicted on the basis of a pQCD calculation with
parton densities that depend on $x$ according to the linear BFKL
(Balitsky--Fadin--Kuraev--Lipatov) \cite{BalitL1,KuraeLF1} evolution
equation.

The counterpart of such a large occupation number is that one can
treat the small-$x$ partons by classical color fields instead of
particles. To that effect, the McLerran-Venugopalan model
\cite{McLerV1,McLerV2,McLerV3} is a hybrid description, in which
the small-$x$ partons are described by classical fields, and where the
large-$x$ partons -- fast and therefore frozen by time dilation -- are
described as static color sources at the origin of the classical
fields, in agreement with the fact that small-$x$ partons are radiated
by bremsstrahlung from the large-$x$ ones. Originally, the MV model
dealt with large nuclei, with a large number of high-$x$ partons (the
number of valence quarks is $3A$ if $A$ is the atomic number of the
nucleus). In the MV model, the large-$x$ color sources are described
by a statistical distribution, which they argued could be taken to be
a Gaussian for a large nucleus at moderately small $x$ (see also
\cite{JeonV1} for a recent discussion of this point).

Since these early days, this model has become an effective theory, the
so-called ``Color Glass Condensate'' (CGC)
\cite{IancuLM1,IancuLM2,FerreILM1}. Since the separation between large
$x$ and small $x$ is arbitrary, no physical quantity should depend on
it. This arbitrariness leads to a renormalization group equation, the
so-called JIMWLK
(Jalilian-Marian--Iancu--McLerran--Weigert--Leonidov--Kovner) equation
\cite{JalilKLW1,JalilKLW2,JalilKLW3,JalilKLW4,KovneM1,KovneMW3,JalilKMW1,IancuLM1,IancuLM2,FerreILM1},
that describes how the statistical distribution of color sources
changes as one moves the separation between large and small $x$. This
functional evolution equation can also be expressed as an infinite
hierarchy of evolution equations for correlators \cite{Balit1}, and
has a quite useful -- and much simpler -- large $N_c$ mean-field
approximation \cite{Kovch3}, known as the Balitsky-Kovchegov equation.

In the collision of two nuclei at high energy, gluon production is
dominated by the classical field approximation, and calculating it
requires to solve the classical Yang-Mills equations for two color
sources -- one for each projectile -- moving at the speed of light in
opposite directions. This problem has been studied numerically in
\cite{KrasnV1,KrasnV2,KrasnNV1,KrasnNV2,Lappi1} for the
boost-invariant case, with extensions to include the rapidity
dependence \cite{RomatV1,RomatV2}. But in fact, for collisions
involving one small projectile -- like proton-nucleus collisions --
one can assume that the color sources that describe this small
projectile are weak and compute the relevant amplitude only at lowest
order in this source. When this is allowed, it is possible to obtain
analytical expressions for amplitudes and cross-sections.  This was
done in a number of approaches for single quark or gluon production
\cite{KovchM3,KovneW1,KovchT1,DumitM1,DumitJ1,DumitJ2,GelisJ3,BlaizGV1,GelisM1,NikolS1},
as well as for quark-antiquark production
\cite{GelisV1,Tuchi1,BlaizGV2,JalilK2,NikolSZ1,NikolSZ2,NikolSZ3,FujiiGV1,BaierKNW1,FujiiGV2}
(see \cite{JalilK1} for a review). In this paper, we are going to
limit our discussion to the case of single gluon production.

One of the main features of the gluon production cross-section in
proton-nucleus obtained in the CGC framework is that it includes all
the multiple scatterings on the sources contained in the nucleus. In
this paper, we discuss the distribution in the number of these
scatterings. In particular, we study how the momentum of a
high-$p_\perp$ final gluon is balanced by the recoiling momenta of the
struck nuclear color sources. This question has practical applications
in discussing whether one could observe a loss of back-to-back
correlations at high $p_\perp$ -- for instance events with a single
high-momentum jet in the final state -- in collisions between a proton
and a saturated nucleus. Another application of our study occurs when
one tries to relate the multiplicity and the impact parameter of the
collision. And of course, one may also try to characterize the nuclear
partonic content at low $x$ from the distribution of ``debris'' that
are produced in the collision with the proton. 

Note that the manifestations of the Color Glass Condensate on the
back-to-back correlations have already been investigated in various
approaches \cite{BaierKNW1,KharzLM2}, by looking at the angular
correlations between pairs of hard particles. This azimuthal
correlation has been measured for deuteron-gold collisions by the STAR
collaboration at RHIC \cite{Adamsa2}, which observed that the pattern
of azimuthal correlations in d--Au collisions is very similar to that
found in pp collisions. In particular, it has a marked peak in the
correlation function at $180$ degrees -- indicating that jets come in
pairs. In the present paper, we address a different question, which
involves most of the same physics: we do not consider the angle of
emission of the particles, but instead we keep track of the number of
recoils above a certain threshold $k_\perp^{\rm min}$ given a
``trigger'' particle with momentum $k_\perp$, and we calculate the
probability for having a given number of such recoils.

Our paper is organized as follows. In section \ref{sec:cs}, we recall
the CGC formula for gluon production in proton-nucleus collisions,
and we also discuss the Glauber interpretation of this formula. In
section \ref{sec:dist}, we show how to calculate the probability $P_n$
of having $n$ scatterings in which the recoil momentum is larger than
a certain threshold $k_\perp^{\rm min}$, when the produced gluon has
acquired the momentum $k_\perp$ in the nucleus. This is done by
constructing a generating function for these probabilities. In section
\ref{sec:MV}, we present numerical results for this distribution in
the MV model, as well as simple analytical calculations that explain
the most salient features; and in section \ref{sec:asympt} we compare
them with what happens in a model that has shadowing and geometrical
scaling. Finally, in section \ref{sec:impact}, we discuss the
possibility of estimating the impact parameter of the collision from
the multiplicity in the final state.

\section{Gluon production in proton-nucleus collisions}
\label{sec:cs}
Here, we use without rederiving it the formula for the gluon yield
obtained in \cite{BlaizGV1} (Eq.~(107)). According to this formula,
the number of gluons produced per unit of transverse momentum and per
unit of rapidity reads:
\begin{equation}
\frac{d\overline{N}_g}{d^2\q_\perp dy}
=\frac{1}{16\pi^3 q_\perp^2}
\int \frac{d^2\k_\perp}{(2\pi)^2}
k_\perp^2 C(\k_\perp)\, \varphi_p(\q_\perp-\k_\perp)\; ,
\label{eq:nbar-pA}
\end{equation}
where $\varphi_p$ is the proton non-integrated gluon distribution,
that we won't need to specify further in the following. The function
$C(k_\perp)$, introduced in \cite{GelisP1}, is the Fourier transform
of a correlator of Wilson lines:\footnote{The function $C(\k_\perp)$ is
thus related to the Fourier transform of the cross-section between a
color dipole and the nucleus. Hence, in the case of a quark-antiquark
dipole, one can use this connection in order to relate proton-nucleus
collisions and Deep Inelastic Scattering on nuclei \cite{GelisJ3}.}
\begin{equation}
C(k_\perp)\equiv
\int d^2\x_\perp e^{i\k_\perp\cdot\x_\perp}\;
\frac{1}{N_c^2-1}\,{\rm Tr}\,
\left<U^\dagger(0)U(\x_\perp)\right>\; .
\end{equation}
$U(\x_\perp)$ is a Wilson line in the adjoint representation of
$SU(N_c)$, evaluated in the color field produced by the sources that
describe the nucleus, and the brackets $\big<\cdots\big>$ denote an
averaging over these color sources. Equation~(\ref{eq:nbar-pA}) is
accurate to the lowest order in the density of color sources contained
in the proton, and to all orders in the color density of the
nucleus. Thus, a way to picture its content is to say that a gluon of
the wave-function of the proton travels through the color field of the
nucleus before being produced.

At first sight, it looks like the process taken into account by
Eq.~(\ref{eq:nbar-pA}) is a $2\to 1$ process, in which one gluon from
the proton (with transverse momentum $\q_\perp-\k_\perp$) merges with a
gluon from the nucleus (with transverse momentum $\k_\perp$) in order
to produce the final gluon of transverse momentum $\q_\perp$. One
might therefore be tempted to conclude that the Color Glass Condensate
predicts the production of {\sl monojets} in proton-nucleus
collisions. However, this conclusion is too simplistic. The first
reason is of course that transverse momentum is conserved in the CGC
framework. This means that if the final gluon has acquired a large
momentum $\k_\perp$ while going through the nucleus, this momentum
must come from the color sources present in the nucleus. In other
words, if one sums the recoil transverse momenta of the sources struck
by the propagating gluon, they must add up to $-\k_\perp$. 

\begin{figure}[htbp]
\begin{center}
\resizebox*{!}{3.5cm}{\includegraphics{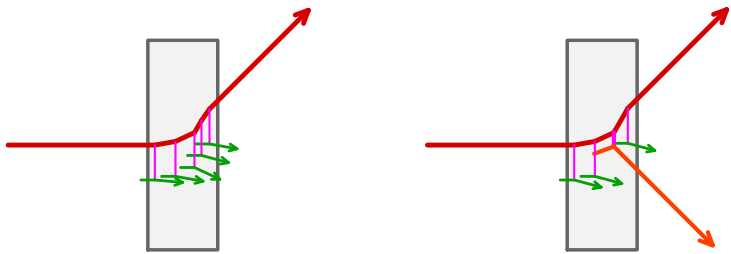}}
\end{center}
\caption{\label{fig:01}Two possible scenarios for the recoiling
scattering centers in the production of a high-$p_\perp$
particle. Left: the large $p_\perp$ of the produced particle is
compensated by many semi-hard recoils. Right: one recoil absorbs
almost all the $p_\perp$.}
\end{figure}

Therefore, the real issue in order to conclude about the possible
existence of monojets is whether the recoil momentum is shared among
many sources (each of them acquiring only a small momentum), or on the
contrary absorbed mostly by a single source (see Figure \ref{fig:01}
for a cartoon illustrating the two situations). If the first scenario
holds, then indeed one would have a high-$q_\perp$ jet whose momentum
is balanced by many soft recoiling particles -- an event topology that
would be close to one's idea of a ``monojet''. In the second scenario,
one would have a pair of high-$q_\perp$ particles, with almost
opposite transverse momenta, in agreement to what perturbative QCD
would predict.

This interpretation in terms of multiple scatterings is particularly
transparent in the case where the distribution of color sources in the
nucleus has only Gaussian correlations. This is the case of the
McLerran-Venugopalan model (in which case the Gaussian distribution is
local), and also of the asymptotic regime believed to be reached after
evolution to large rapidities with the JIMWLK evolution equation (in
which case it is a non-local Gaussian distribution)
\cite{IancuIM2}. Indeed, for a Gaussian distribution of nuclear color
sources, it is possible to rewrite the function $C(k_\perp)$ in a form
that has an obvious Glauber interpretation.  Let us reproduce here the
main result of the appendix C of \cite{BlaizGV1}. Following
Eqs.~(C.5-6) of this reference, we can rewrite the function
$C(k_\perp)$ as follows
\begin{eqnarray}
&&C(k_\perp)=e^{-\mu_0^2 \sigma_{\rm tot}}
\sum_{n=0}^{+\infty}
\rho^n \int\limits_0^L dz_1
\int\limits_{z_1}^L dz_2\cdots
\int\limits_{z_{n-1}}^L dz_n
\int
\frac{d^2\k_{1\perp}}{(2\pi)^2}\cdots\frac{d^2\k_{n\perp}}{(2\pi)^2}
\nonumber\\
&&\qquad\qquad\qquad
\times(2\pi)^2\delta(\k_{1\perp}+\cdots+\k_{n\perp}-\k_\perp)
\sigma(\k_{1\perp})\cdots\sigma(\k_{n\perp})\; .
\label{eq:glauber}
\end{eqnarray}
In this formula, $\rho$ is the number of scattering centers per unit
of volume of the nucleus (assumed to be uniform), $L$ is the
longitudinal size of the nucleus, $\sigma(\k_\perp)$ is the
differential cross-section of a gluon with a scattering center of the
nucleus, and $\sigma_{\rm tot}$ is the integral of the latter over
$\k_\perp$. Finally, $\mu_0^2\equiv \rho L$ is the density of
scattering centers per unit of transverse area.

\section{Distribution of struck scattering centers}
\label{sec:dist}
\subsection{Definition}
In Eq.~(\ref{eq:glauber}), the index $n$ is the number of collisions
of the gluon coming from the proton while it travels through the
nucleus, and the exponential in the prefactor serves to unitarize the
overall sum. Note that the integral over $\k_\perp$ of the function
$C(\k_\perp)$ is equal to one, which means that this function should
be interpreted as the probability for the gluon to acquire the
momentum $\k_\perp$ while going through the nucleus. The term of order
$n$ in this formula is therefore the probability that the gluon be
deflected by a transverse momentum $\k_\perp$ and undergo exactly $n$
scatterings. By dividing this term by $C(\k_\perp)$, we obtain the
conditional probability that a gluon that comes out with a momentum
$\k_\perp$ has scattered $n$ times:
\begin{eqnarray}
&&P_n(\k_\perp)
=
\frac{e^{-\mu_0^2 \sigma_{\rm tot}}}{C(\k_\perp)}
\rho^n \int\limits_0^L dz_1
\int\limits_{z_1}^L dz_2\cdots
\int\limits_{z_{n-1}}^L dz_n
\int
\frac{d^2\k_{1\perp}}{(2\pi)^2}\cdots\frac{d^2\k_{n\perp}}{(2\pi)^2}
\nonumber\\
&&\qquad\qquad\qquad
\times(2\pi)^2\delta(\k_{1\perp}+\cdots+\k_{n\perp}-\k_\perp)
\sigma(\k_{1\perp})\cdots\sigma(\k_{n\perp})\; .
\label{eq:Pn}
\end{eqnarray}

So far, we have been a bit sloppy regarding the infrared behavior of
the integrals over the transverse momenta that appear in
Eqs.~(\ref{eq:glauber}) and (\ref{eq:Pn}). However, in the MV model
for instance, $\sigma(\k_\perp)$ behaves as $k_\perp^{-4}$ at small
$k_\perp$ and it is necessary to introduce an infrared cutoff
$\Lambda$ in order for the integrals to be finite. It is well known
that, although each integral behave as $\Lambda^{-2}$, a partial
cancellation occurs with the prefactor $\exp(-\mu_0^2 \sigma_{\rm
tot})$ so that $C(\k_\perp)$ is only logarithmically sensitive to this
cutoff\footnote{In the individual probabilities $P_n$ however, this
cancellation does not occur and one has a quadratic sensitivity to
the cutoff $\Lambda$.}. Physically, this cutoff emerges from color
neutralization that occurs on distance scales of the order of the
nucleon size. Therefore, one should take $\Lambda\approx\Lambda_{_{\rm
QCD}}$.

This cutoff is of course also necessary in order to define the
probabilities $P_n$, so that they should in fact be
interpreted as probabilities to have $n$ scatterings with a momentum
transfer larger than $\Lambda$. In the case of the $P_n$'s, we can
even push this logic further by defining the probabilities to have $n$
scatterings with a momentum transfer larger than a certain
$k_\perp^{\rm min}$ which is not necessarily related to $\Lambda$, and
an arbitrary number of scatterings with a momentum transfer between
$\Lambda$ and $k_\perp^{\rm min}$. By doing so, we can explore how the
distribution of the number of scatterings evolves with their
``hardness''. Let us denote $P_n(k_\perp|k_\perp^{\rm min})$ this
probability. It is very easy to extract the relevant piece from
Glauber formula, Eq.~(\ref{eq:glauber}):
\begin{eqnarray}
&&P_n(\k_\perp|k_\perp^{\rm min})
=
\frac{e^{-\mu_0^2 \sigma_{\rm tot}}}{C(\k_\perp)}
\sum_{p=0}^{+\infty}
\rho^{p+n} \int\limits_0^L dz_1
\int\limits_{z_1}^L dz_2\cdots
\int\limits_{z_{p+n-1}}^L dz_{p+n}
\nonumber\\
&&
\qquad\qquad
\times
\int_{\Lambda}^{k_\perp^{\rm min}}
\frac{d^2\k_{1\perp}}{(2\pi)^2}\cdots\frac{d^2\k_{p\perp}}{(2\pi)^2}
\int_{k_\perp^{\rm min}}
\frac{d^2\k_{p+1\perp}}{(2\pi)^2}\cdots\frac{d^2\k_{p+n\perp}}{(2\pi)^2}
\nonumber\\
&&\qquad\qquad
\times(2\pi)^2\delta(\k_{1\perp}+\cdots+\k_{p+n\perp}-\k_\perp)
\sigma(\k_{1\perp})\cdots\sigma(\k_{p+n\perp})\; .
\label{eq:Pn1}
\end{eqnarray}
In this formula, $n$ is the number of scatterings with momentum
transfer larger than $k_\perp^{\rm min}$ and $p$ the number of
scatterings with momentum transfer between $\Lambda$ and $k_\perp^{\rm
min}$.

\subsection{Generating function}
Although a direct numerical evaluation of Eq.~(\ref{eq:Pn1}) is in
principle feasible, it turns out to be easier to compute the following
generating function instead:
\begin{equation}
F(z,k_\perp|k_\perp^{\rm min})
\equiv
\sum_{n=0}^{+\infty}P_n(k_\perp|k_\perp^{\rm min})
\;z^n\; .
\label{eq:F-def}
\end{equation}
From this function, it is straightforward to go back to the
probabilities $P_n$ by the following formula:\footnote{Another approach
  to obtain the probabilities from the generating function is to
  compute the successive derivatives of the generating function at
  $z=0$. However, this would require to evaluate derivatives of high
  order, which is very difficult to do numerically.}
\begin{equation}
P_n(k_\perp|k_\perp^{\rm min})
=
\int\limits_0^{2\pi}
\frac{d\theta}{2\pi}\;
 e^{-in\theta}\;F(e^{i\theta},k_\perp|k_\perp^{\rm min})\; .
\end{equation}
Therefore, it will be sufficient to calculate the generating function
for complex $z$'s on the unit circle. In practice, one should evaluate
the generating function for a finite number (usually a power of two)
of values $z=e^{i\theta}$, with the angles $\theta$ equally spaced on
the circle, and then evaluate the Fourier sum by the fast 
Fourier-transform algorithm.

It is easy to replace $P_n$ by its expression in Eq.~(\ref{eq:F-def}),
and to perform the sum explicitly. In order to disentangle the various
variables $\k_{i\perp}$, one must replace the delta function by its
Fourier representation. This leads to:
\begin{eqnarray}
&&
F(z,k_\perp|k_\perp^{\rm min})
=
\frac{1}{C(\k_\perp)}
\int d^2\x_\perp e^{-i\k_\perp\cdot \x_\perp}
\nonumber\\
&&
\times
\exp\left\{\mu_0^2\left[
\int\limits_\Lambda^{k_\perp^{\rm min}}
\!\!\frac{d^2\l_\perp}{(2\pi)^2}
(e^{i\l_\perp\cdot\x_\perp}\!-\!1)\sigma(\l_\perp)
+
\!\!\!\int\limits_{k_\perp^{\rm min}}
\!\!\frac{d^2\l_\perp}{(2\pi)^2}
(ze^{i\l_\perp\cdot\x_\perp}\!-\!1)\sigma(\l_\perp)
\right]\right\}\; .
\nonumber\\
&&
\label{eq:F2}
\end{eqnarray}
Note that, for $z=1$, the numerator of this formula is identical to
$C(\k_\perp)$. This was of course expected, since
$F(1,k_\perp|k_\perp^{\rm min})=1$ (because this is the sum of all the
probabilities $P_n$). As one can see, the only difference between the
calculation of $C(\k_\perp)$ and of the numerator in Eq.~(\ref{eq:F2})
is that the exponential $\exp(i\l_\perp\cdot\x_\perp)$ is weighted by
a factor $z$ for the values of $l_\perp$ above $k_\perp^{\rm
min}$. Therefore, calculating the generating function can be done via
a fairly minor modification\footnote{This observation also indicates
how to construct the generating function for probabilities that are
more general than the ones considered here: in order to compute the
probabilities $P_n({\bs\Omega})$ to produce $n$ particles in some
region ${\bs\Omega}$ of the single particle phase-space, one must
weight the exponential $\exp(i\l_\perp\cdot\x_\perp)$ by a factor $z$
when $\l_\perp\in{\bs\Omega}$. This approach could be used in order to
study the recoils in a specific angular sector for instance.} of the
numerical methods used in order to calculate $C(\k_\perp)$.

In fact, as one can readily see, in order to calculate the argument of
the exponential in Eq.~(\ref{eq:F2}), it is sufficient to compute the
following two integrals,
\begin{eqnarray}
&&
A(x_\perp)
\equiv
\int\limits_{\Lambda}\frac{d^2\l_\perp}{(2\pi)^2}
(e^{i\l_\perp\cdot\x_\perp}\!-\!1)\,\sigma(\l_\perp)
\; ,
\nonumber\\
&&
B(x_\perp|k_\perp^{\rm min})
\equiv
\int\limits_{k_\perp^{\rm min}}\frac{d^2\l_\perp}{(2\pi)^2}
e^{i\l_\perp\cdot\x_\perp}\sigma(\l_\perp)
\; ,
\label{eq:AB}
\end{eqnarray}
as a function of $\x_\perp$ and $k_\perp^{\rm min}$. 

In actual numerical calculations, the lower limits, at
$l_\perp=\Lambda$ in $A$ and at $l_\perp=k_\perp^{\rm min}$ in $B$,
are implemented by multiplying the integrand respectively by
$\vartheta(l_\perp/\Lambda)$ and $\vartheta(l_\perp/k_\perp^{\rm
min})$. The function $\vartheta(x)$ interpolates between $0$ at small
$x$ and $1$ at large $x$, the transition between the two regimes being
located around $x=1$. One could in principle take for $\vartheta(x)$
the ordinary step function $\theta(x)$, which corresponds to sharp
lower limits -- as written in eqs.~(\ref{eq:AB}) -- but such a choice
generally leads to an oscillatory behavior of the functions
$A(x_\perp)$ and $B(x_\perp|k_\perp^{\rm min})$ as a function of
$x_\perp$. Choosing a function $\vartheta(x)$ that has a smooth
transition between $0$ and $1$ is helpful in order to tame these
oscillations.

Once the integrals $A$ and $B$ have been calculated, one can write:
\begin{equation}
C(\k_\perp)=
\int d^2\x_\perp\; e^{-i\k_\perp\cdot\x_\perp}\;
e^{\mu_0^2\, A(\x_\perp)}\; ,
\label{eq:C3}
\end{equation}
and then
\begin{equation}
F(z,k_\perp|k_\perp^{\rm min})
=
\frac{1}{C(\k_\perp)}
\int d^2\x_\perp\; e^{-i\k_\perp\cdot \x_\perp}\;
e^{\mu_0^2\,\left[ A(\x_\perp)+(z-1)B(\x_\perp|k_\perp^{\rm min})\right]}\; .
\label{eq:F3}
\end{equation}

\subsection{Models for $\sigma(\k_\perp)$}
In the rest of this paper, we consider two different models for
the differential cross-section $\sigma(\l_\perp)$.

The first of these two models is the McLerran-Venugopalan model
\cite{McLerV1,McLerV2,McLerV3}, which assumes a local Gaussian
distribution of color charges in the transverse plane for the
nucleus. It is well known that this leads to\footnote{Here, the
formula has been written in the adjoint representation, since it is a
gluon that propagates through the nucleus. For a quark, one would
simply have to replace the color factor $N_c$ by $C_{\rm f}\equiv
(N_c^2-1)/2N_c$.}
\begin{equation}
\sigma(\l_\perp)=
\frac{1}{2}\frac{g^4 N_c}{l_\perp^4}\; .
\end{equation}
In the MV model, one can have important rescattering effects (tuned
via the density parameter $\mu_0^2$), but there is no leading-twist
shadowing. Note that the saturation momentum $Q_s$ is given by:
\begin{equation}
Q_s^2=\frac{ g^4 C_{\rm f}}{4\pi} \mu_0^2 
\ln\left(\frac{\mu_0^2}{\Lambda_{_{QCD}}^2}\right)\; .
\end{equation}
Here, we have written the saturation momentum in the fundamental
representation, in order to facilitate the comparison with the values
of $Q_s$ extracted from Deep Inelastic Scattering at HERA.

The second model we will consider is based on a Gaussian effective
theory that describes the gluonic content of a nucleus evolved to very
small values of $x$, discussed in \cite{IancuIM2}. It corresponds to
the choice
\begin{equation}
\mu_0^2\sigma(l_\perp)=
\frac{2\pi}{\gamma c}
\,
\frac{Q_s^2}{l_\perp^2}
\ln\left(
1+\Big(\frac{Q_s^2}{l_\perp^2}\Big)^{\gamma}
\right)\; .
\label{eq:IIM}
\end{equation}
In this model, hereafter referred to as the ``asymptotic model'',
$c\approx 4.84$ and $\gamma$ is an anomalous dimension whose value is
$\gamma\approx 0.64$. One of the peculiarities of this model is that
it has the property of ``geometrical scaling'', since it depends on
the momentum $l_\perp$ and on $Q_s$ only via the ratio
$l_\perp/Q_s$. Contrary to the MV model, this non-local Gaussian model
has significant leading-twist shadowing, whose strength is
controlled by the anomalous dimension $\gamma$ (more precisely by the
departure of $\gamma$ from 1).

\section{Results in the MV model}
\label{sec:MV}
\subsection{Multiplicity distribution}
\begin{figure}[htbp]
\begin{center}
\resizebox*{9cm}{!}{\rotatebox{-90}{\includegraphics{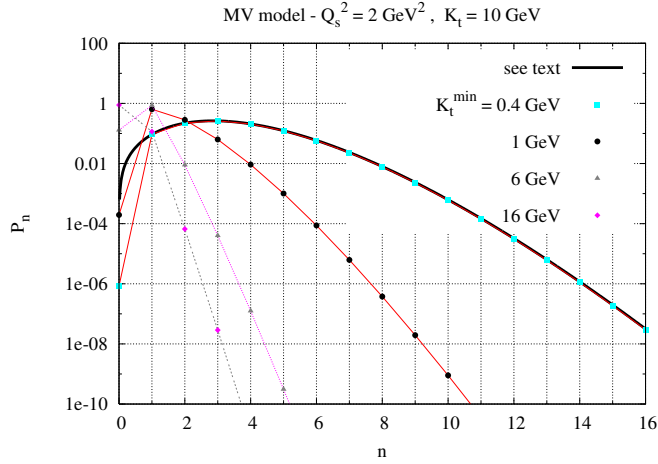}}}
\end{center}
\caption{\label{fig:Pn}Distribution of the probabilities $P_n$ in the
MV model. The produced particle has acquired a transverse momentum
$k_\perp = 10$~GeV and the saturation momentum is set to the value
$Q_s^2=2$~GeV${}^2$. The threshold $k_\perp^{\rm min}$ for counting
the recoiling particles takes values between $0.4$ and $16$~GeV. The
black solid line represents the result of an approximate calculation
(see text).}
\end{figure}

Let us first start by displaying some results in the MV model. In
Figure \ref{fig:Pn}, we first show the distribution of the
probabilities $P_n$ as a function of $n$, for $Q_s^2=2$~GeV${}^2$ and
various values of the threshold momentum $k_\perp^{\rm min}$. 
One can see that the width of the multiplicity distribution decreases
with an increasing $k_\perp^{\rm min}$. This is of course quite
natural, since by increasing $k_\perp^{\rm min}$ it becomes less and
less likely to have events in which there are a large number of
recoils. Note also that for all $k_\perp^{\rm min}$ such that $Q_s\ll
k_\perp^{\rm min}\lesssim k_\perp$, the most likely number of recoils
is $n=1$, while for $k_\perp < k_\perp^{\rm min}$ the most likely
situation is $n=0$.

We can in fact understand analytically this distribution $P_n$ in the
situation where the momentum exchange $k_\perp$ between the incoming
gluon and the nucleus is much larger than the other scales,
$Q_s,k_\perp^{\rm min}\ll k_\perp$. This means that we need only to
estimate the functions $A$ and $B$ defined in Eq.~(\ref{eq:AB}) for
values of $x_\perp$ that are much smaller than the inverse saturation
momentum, $x_\perp\ll Q_s^{-1}$, and much smaller than $(k_\perp^{\rm
min})^{-1}$. This allows us to expand the exponential
$\exp(i\l_\perp\cdot \x_\perp)$ in order to evaluate the integral over
$\l_\perp$, leading to the following approximations:
\begin{eqnarray}
&&
A(x_\perp)
\approx 
-\frac{g^4 N_c}{16\pi}\,x_\perp^2\,
\ln\left(\frac{1}{\Lambda x_\perp}\right)\; ,
\nonumber\\
&&
B(\x_\perp|k_\perp^{\rm min})
\approx
\frac{g^4 N_c}{8\pi}
\left[
\frac{1}{k_\perp^{\rm min\ 2}}
-\frac{x_\perp^2}{2}\,
\ln\left(\frac{1}{k_\perp^{\rm min} x_\perp}\right)
\right]\; .
\label{eq:ABapprox}
\end{eqnarray}
Then, in order to evaluate the generating function via
Eqs.~(\ref{eq:C3}) and (\ref{eq:F3}), one can use the following
result, valid at large $k_\perp$,
\begin{equation}
\int d^2\x_\perp\;
e^{-i\k_\perp\cdot\x_\perp}
\;
e^{-C x_\perp^2\ln(x_\perp^0/x_\perp)}
\approx
\frac{8\pi C}{k_\perp^4}\; .
\end{equation}
(The value of the constant $x^0_\perp$ has no influence on this result
in the limit of large $k_\perp$.)  Thanks to this formula, we obtain
immediately
\begin{eqnarray}
F(z,k_\perp|k_\perp^{\rm min})
\approx
z\;e^{\frac{g^4 N_c \mu_0^2}{8\pi (k_\perp^{\rm min})^2}(z-1)}\; .
\label{eq:F4}
\end{eqnarray}
One can see that in this limit, the generating function is universal
in the sense that it does not depend on the momentum $k_\perp$
acquired by the incoming gluon. Moreover, the probability of having
zero scatterings with a recoil above $k_\perp^{\rm min}$,
$P_0=F(z=0,k_\perp|k_\perp^{\rm min})$, is zero. In other words, when
$k_\perp^{\rm min}\ll k_\perp$, there must be at least one scattering
above $k_\perp^{\rm min}$ in order to give such a large $k_\perp$ to
the incoming gluon.

We can go a bit further, since it is easy to recognize that the
generating function obtained in Eq.~(\ref{eq:F4}) corresponds to the
following distribution of probabilities:
\begin{eqnarray}
&&
P_0=0\; ,
\nonumber\\
&&
P_n= \frac{\overline{n}^{\,n-1}}{(n-1)!}\; e^{-\overline{n}}\quad\mbox{with\ \ }
\overline{n}\equiv \frac{g^4 N_c \mu_0^2}{8\pi (k_\perp^{\rm min})^2}\; .
\label{eq:Pn3}
\end{eqnarray}
In other words, the distribution of multiplicities is a Poisson
distribution shifted by one unit. The physical meaning of this shift
will become transparent later in the discussion. In Figure
\ref{fig:Pn}, we have compared for $k_\perp^{\rm min}=0.4$~GeV the
numerically evaluated $P_n$'s with such a shifted Poisson
distribution, and as one can see the two agree extremely well (except
for $P_0$, which is very small but not exactly zero). 

Note however that the value of $\overline{n}$ we had to use in this
fit differs by about 25\% from the predicted value given in
Eq.~(\ref{eq:Pn3}). This kind of deviation is expected, because this
formula for $\overline{n}$ is only valid for $k_\perp^{\rm min}\gg
\Lambda$, a condition which is at best marginally satisfied for
$k_\perp^{\rm min}=0.4~$GeV (we have taken the infrared cutoff to be
$\Lambda=0.2~$GeV). Moreover, from the approximations that have been
used in order to obtain Eq.~(\ref{eq:ABapprox}), a generating function
of the form $F(z)=z\;\exp(\overline{n}(z-1))$ -- that leads to a
shifted Poisson distribution -- is obtained as long as $k_\perp\gg
Q_s, k_\perp^{\rm min}$. It is only the accurate prediction of the
value of $\overline{n}$ that requires in addition $k_\perp^{\rm
min}\gg \Lambda$. This explains why, despite the fact that
$\overline{n}$ was not very accurately predicted at $k_\perp^{\rm
min}=0.4~$GeV, the obtained distribution was nevertheless of the form
given in Eq.~(\ref{eq:Pn3}) with a very good accuracy, because
$k_\perp=10~$GeV.

\subsection{Number of recoils}
\begin{figure}[htbp]
\vspace{-4mm}
\begin{center}
\resizebox*{9cm}{!}{\rotatebox{-90}{\includegraphics{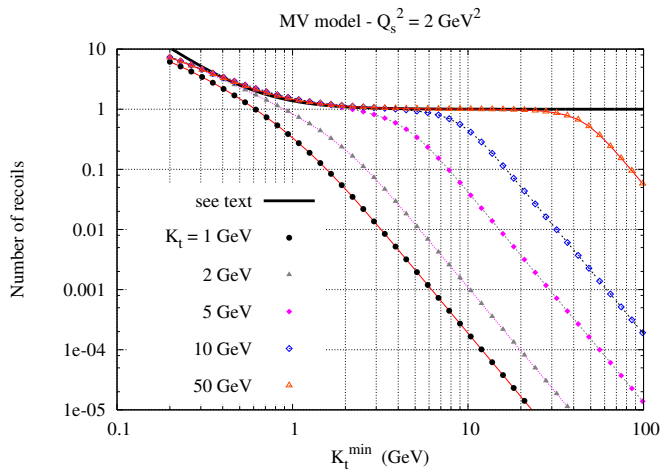}}}
\end{center}
\caption{\label{fig:Nbar-mv}Number of recoiling scattering centers in the
MV model, as a function of the threshold $k_\perp^{\rm min}$. The
saturation momentum is set to the value $Q_s^2=2$~GeV${}^2$, and the
momentum $k_\perp$ of the produced particle is varied between $1$ and
$50$~GeV.}
\end{figure}
Next, we display in Figure \ref{fig:Nbar-mv} the average number of
recoils above the threshold $k_\perp^{\rm min}$, defined as
\begin{equation}
N(k_\perp|k_\perp^{\rm min})
\equiv
\sum_{n=1}^\infty
n\,P_n(k_\perp|k_\perp^{\rm min})\; ,
\label{eq:N4}
\end{equation}
 as a function of $k_\perp^{\rm min}$, for various momenta $k_\perp$
and a fixed $Q_s^2=2$~GeV${}^2$. We see that the number of recoils
grows significantly at small $k_\perp^{\rm min}$, and tends in this
region to become universal and independent of $k_\perp$. Moreover, a
striking feature of this number of recoils is that it is very close to
unity for any value of $k_\perp^{\rm min}$ such that $Q_s\ll
k_\perp^{\rm min}\lesssim k_\perp$. This means that when the gluon
acquires a large momentum $k_\perp$ from the nucleus, there is always
one hard recoil -- and only one -- that provides most of this large
momentum.

Again, it is possible to have an analytic understanding of these
properties of the average number of recoils from
Eq.~(\ref{eq:F4}). Indeed, the average number of recoils is given by
the derivative of the generating function at $z=1$, and we obtain
\begin{equation}
N(k_\perp|k_\perp^{\rm min})
\approx
1
+
\frac{g^4 N_c \mu_0^2}{8\pi (k_\perp^{\rm min})^2}\; .
\label{eq:N4-1}
\end{equation}
This analytic expression is also displayed in Figure
\ref{fig:Nbar-mv}, and it reproduces well the numerical calculation
for $k_\perp^{\rm min}\lesssim k_\perp$. It deviates from it at very
small $k_\perp^{\rm min}$ due to a non-trivial interplay between
$k_\perp^{\rm min}$ and the infrared cutoff $\Lambda$, which is not
correctly captured by our simple analytic calculation. And of course
this analytical result does not work for $k_\perp^{\rm min}\ge
k_\perp$ because this is outside the range of validity of our
approximations.

\subsection{Momentum distribution of the recoils}
\begin{figure}[htbp]
\vspace{-4mm}
\begin{center}
\resizebox*{9cm}{!}{\rotatebox{-90}{\includegraphics{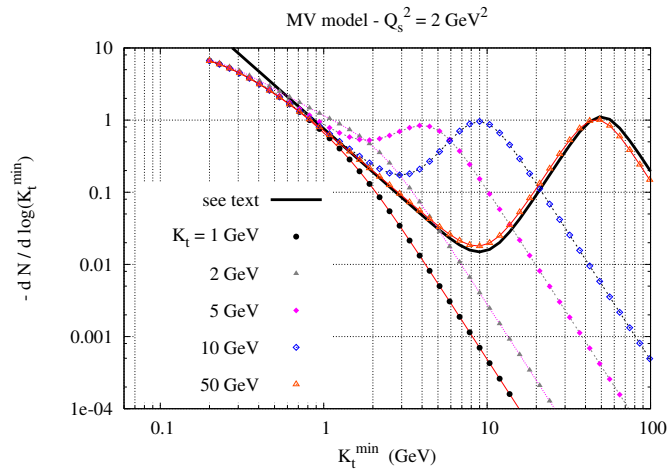}}}
\end{center}
\caption{\label{fig:dNbar-mv}Distribution of the number of recoiling
scattering centers in the MV model, as a function of the threshold
$k_\perp^{\rm min}$. The saturation momentum is set to the value
$Q_s^2=2$~GeV${}^2$, and the momentum $k_\perp$ of the produced
particle is varied between $1$ and $50$~GeV.}
\end{figure}
In Figure \ref{fig:dNbar-mv}, we have taken the derivative of the
average number of recoils with respect to $k_\perp^{\rm min}$, in
order to obtain the momentum distribution of these recoils. At large
$k_\perp$, one can clearly see that this distribution consists of two
components: a universal (almost independent of $k_\perp$) semi-hard
component made of recoils with momenta of the order of $Q_s$ or
smaller, and a component peaked around $k_\perp^{\rm
min}=k_\perp$. The latter peak has an area unity, and it is simply
translated when $k_\perp$ is changed. By taking the derivative of
Eq.~(\ref{eq:N4-1}), one can readily obtain a contribution that
reproduces well the numerical result in the semi-hard region
\begin{equation}
-\frac{dN}{d\ln(k_\perp^{\rm min})}
\approx
\frac{g^4 N_c \mu_0^2}{4\pi (k_\perp^{\rm min})^2}\; .
\label{eq:dN4}
\end{equation}
In fact, it turns out that it is also possible to estimate this
derivative in the region where $k_\perp^{\rm min}$ is comparable to or
larger than $k_\perp$ (both of them being very large compared to
$Q_s$). When both $k_\perp^{\rm min}$ and $k_\perp$ are large compared
to $Q_s$ (i.e. to $\mu_0^2$), it is enough to expand the exponentials
of $A$ and $B$ in Eq.~(\ref{eq:F3}) to first order, and
write\footnote{Note that the $1$ in the Taylor expansion of the
exponential does not contribute at large $k_\perp$ since it only gives
a term proportional to $\delta(\k_\perp)$.}
\begin{equation}
F(z,k_\perp|k_\perp^{\rm min})\approx
\frac
{\int d^2\x_\perp\;e^{-i\k_\perp\cdot\x_\perp}\;\Big[A(x_\perp)+(z-1)B(x_\perp|k_\perp^{\rm min})\Big]}
{\int d^2\x_\perp\;e^{-i\k_\perp\cdot\x_\perp}\;A(x_\perp)}
\; .
\label{eq:F-LT}
\end{equation}
The multiplicity being the derivative of $F$ at $z=1$, we have
\begin{equation}
N(k_\perp|k_\perp^{\rm min})
\approx
\frac
{\int d^2\x_\perp\;e^{-i\k_\perp\cdot\x_\perp}\;B(x_\perp|k_\perp^{\rm min})}
{\int d^2\x_\perp\;e^{-i\k_\perp\cdot\x_\perp}\;A(x_\perp)}
\; .
\end{equation}
Going back to the form (\ref{eq:AB}) of $A$ and $B$, we see that the
integration over $\x_\perp$ simply produces a
$\delta(\k_\perp-\l_\perp)$, making the integral over $\l_\perp$
trivial as well. In this kinematical region, we obtain an extremely
simple result:
\begin{equation}
N(k_\perp|k_\perp^{\rm min})
\approx
\frac{\sigma(\k_\perp)\vartheta(k_\perp/k_\perp^{\rm min})}{\sigma(\k_\perp)}
=\vartheta(k_\perp/k_\perp^{\rm min})\; .
\end{equation}
We see that this component of the multiplicity is nothing but the
cutoff function that we are using in order to separate the momenta
that are below $k_\perp^{\rm min}$ from those that are above.
Therefore, the precise shape of the average number of scatterings for
$k_\perp^{\rm min}$ above $k_\perp$ is not a property of QCD, but
merely reflects the fact that we have an extended rather than a sharp
cutoff. Nevertheless, the interpretation of this contribution is quite
straightforward: when $k_\perp^{\rm min}$ is smaller than $k_\perp$
there is one recoil (that absorbs most of the momentum $k_\perp$), but
it is unlikely that there is a recoil with a momentum bigger than the
momentum $k_\perp$. Taking a derivative with respect to $k_\perp^{\rm
min}$, we obtain the corresponding contribution to the momentum
distribution of the recoils:
\begin{equation}
-\frac{dN}{d\ln(k_\perp^{\rm min})}
\approx
\frac{k_\perp}{k_\perp^{\rm min}}
\;
\vartheta^\prime(k_\perp/k_\perp^{\rm min})\; .
\label{eq:dN5}
\end{equation}
In Figure \ref{fig:dNbar-mv}, we have represented for $k_\perp=50$~GeV
the sum of the contributions given in Eqs.~(\ref{eq:dN4}) and
(\ref{eq:dN5}) (taking for the latter the same ``step function''
$\vartheta(x)$ as the one used in the numerical calculation of the
integral $B$). The sum of these two components reproduces with a
fairly good accuracy the numerical results for all $k_\perp^{\rm min}$
down to $k_\perp^{\rm min}\sim 500$~MeV. The small discrepancy between
our analytical estimate of the peaked contribution and its numerical
value is due to rescattering corrections -- indeed, our derivation of
Eq.~(\ref{eq:dN5}) retains only the leading-twist contribution. As one
can see, the numerically obtained peak is slightly shifted to the left
of the analytical result. This is easy to understand: since there are
a few semi-hard scatterings in addition to the hard one, the hard
scattering needs to provide a little less than the momentum $k_\perp$
acquired by the gluon. This shift is a form of collisional energy loss
(for a cold nuclear medium).

Note that, when the ``step function'' $\vartheta(x)$ becomes a real
step function $\theta(x)$, Eq.~(\ref{eq:dN5}) would imply a peak
proportional to $\delta(k_\perp^{\rm min}-k_\perp)$. However, we
expect that higher-twist corrections to Eq.~(\ref{eq:F-LT}) would be
important in this limit, and they are likely to smear out slightly
the delta peak.

Before considering the $Q_s$ dependence, let us come back to the
Poisson distribution shifted by one unit found in Eq.~(\ref{eq:Pn3}),
when $k_\perp^{\rm min}\ll k_\perp$. The shift by one unit is due to
the fact that, when the threshold $k_\perp^{\rm min}$ is so low
compared to $k_\perp$, there is always at least one scattering
(moreover, we know now that this scattering has a recoil momentum
which is close to $k_\perp$). The meaning of Eq.~(\ref{eq:Pn3}) is
therefore that the remaining $n-1$ -- semi-hard -- scatterings that
come along with this hard scattering have a Poissonian distribution,
which merely reflects the fact that they are independent from one
another.

\subsection{Dependence on $Q_s$}
\begin{figure}[htbp]
\vspace{-2mm}
\begin{center}
\resizebox*{9cm}{!}{\rotatebox{-90}{\includegraphics{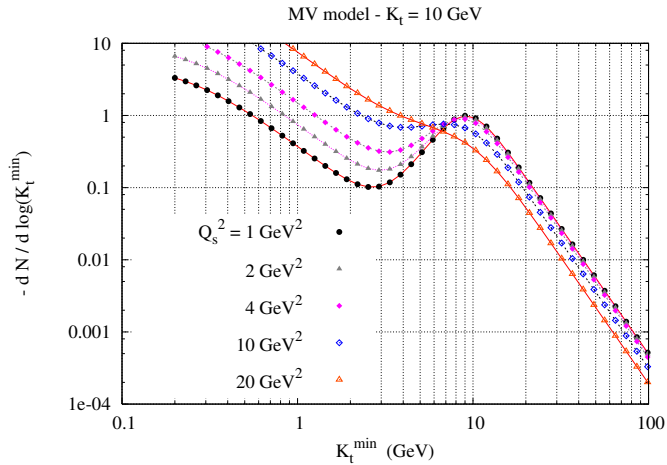}}}
\end{center}
\caption{\label{fig:dNbar-mv-qs}$Q_s$ dependence of the distribution of
the number of recoiling scattering centers in the MV model. The
saturation momentum is varied between $Q_s^2=1$~GeV${}^2$ and
$Q_s^2=20$~GeV${}^2$, and the momentum $k_\perp$ of the produced
particle is set to $10$~GeV.}
\vspace{-2mm}
\end{figure}

Finally, let us have a look at the dependence on the saturation
momentum. For this, we set the momentum $k_\perp$ acquired by the
gluon to $10$~GeV, and we study the momentum distribution of the
recoils for various values of $Q_s^2$. The results of this analysis
are displayed in Figure \ref{fig:dNbar-mv-qs}.
As long as the saturation scale $Q_s$ remains small compared to
$k_\perp$, only the semi-hard part of the distribution is affected by
changes of $Q_s$, while the peak around $k_\perp^{\rm min}=k_\perp$
remains unchanged. The latter result is due to the fact that, since
this peak is well approximated by a leading-twist calculation, it must
be independent of saturation physics with the same accuracy. It is
only when $Q_s$ becomes very large that one cannot neglect higher-twist 
corrections at large $k_\perp^{\rm min}$, and that the peak at
$k_\perp^{\rm min}=k_\perp$ eventually disappears. Since the distribution
of semi-hard recoils is quite sensitive to the value of $Q_s$ (it is
proportional to $g^4\mu_0^2$, which is proportional to $Q_s^2$ up to a
logarithm), it could perhaps be used as a way to estimate $Q_s$.

The disappearance of the peak also provides a qualitative answer to
our initial question regarding the possible existence of monojets: any
parton produced with a $k_\perp$ which is much larger than the
saturation momentum in the nucleus must have its momentum balanced by
another parton on the opposite side (the latter comes from the
scattering center that has undergone the hard collision). But all the
partons with a transverse momentum comparable to or smaller than $Q_s$
need not have their momentum balanced by a leading parton on the
opposite side, since it can be balanced by several softer particles
(coming from the semi-hard component of the distribution of
recoils). As long as the saturation momentum remains relatively small,
say $Q_s\sim 1-3$~GeV, this conclusion is not going to alter one's
common expectations regarding jets: all hard jets with a momentum
larger than say $10$~GeV must come in pairs. It is only for a very
large $Q_s$ that one would start seeing non-conventional event
topologies where a hard jet would have its momentum balanced by a
large number of softer particles.

\section{Effect of leading-twist shadowing}
\label{sec:asympt}
\begin{figure}[htb!]
\vspace{-4mm}
\begin{center}
\resizebox*{9cm}{!}{\rotatebox{-90}{\includegraphics{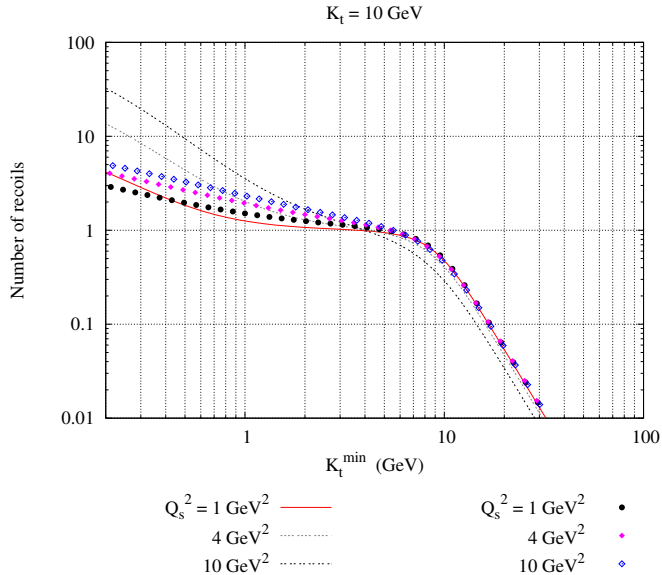}}}
\end{center}
\vspace{-2mm}
\caption{\label{fig:Nbar-mv-as-qs}$Q_s$ dependence of the number of
recoiling scattering centers in the asymptotic model, compared to the
MV model. The saturation momentum is varied between
$Q_s^2=1$~GeV${}^2$ and $Q_s^2=10$~GeV${}^2$, and the momentum
$k_\perp$ of the produced particle is set to $10$~GeV. Lines: MV
model. Dots: asymptotic model.}
\end{figure}
Let us now briefly compare the results previously obtained using the
MV model, with those one obtains by using the model defined by
Eq.~(\ref{eq:IIM}). Basically, the two models -- at an identical $Q_s$
-- differ by the nature of the correlations among the color charges in
the nucleus. In particular, the MV model does not have any leading-twist 
shadowing, while the second model has an anomalous dimension
$\gamma$ different from unity and thus provides some shadowing. It is
believed that the latter model is a better description of a nucleus at
very small momentum fractions $x$.
\begin{figure}[htbp]
\begin{center}
\resizebox*{9cm}{!}{\rotatebox{-90}{\includegraphics{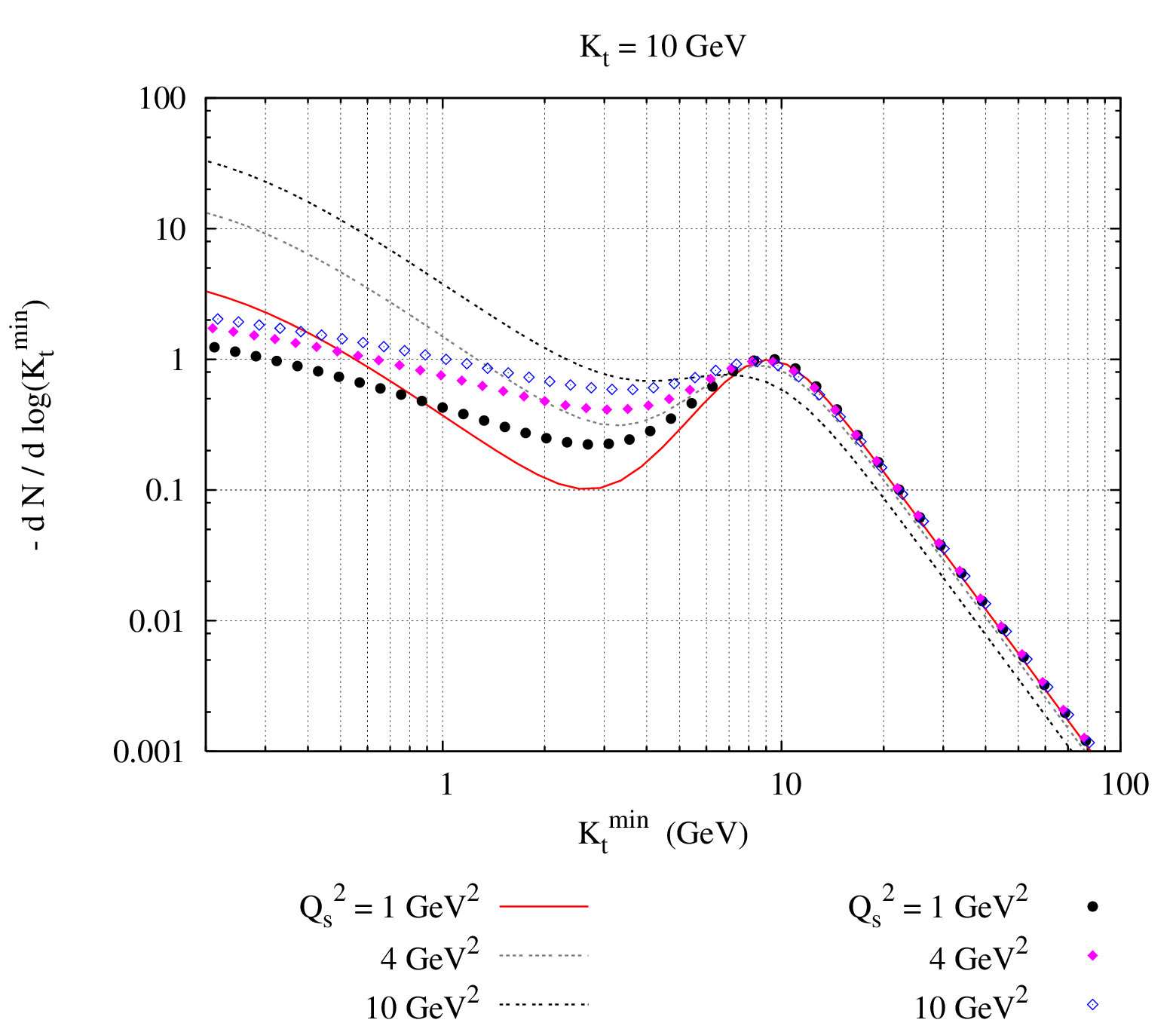}}}
\end{center}
\vspace{-2mm}
\caption{\label{fig:dNbar-mv-as-qs}$Q_s$ dependence of the
distribution of the number of recoiling scattering centers in the
asymptotic model, compared to the MV model. The saturation momentum is
varied between $Q_s^2=1$~GeV${}^2$ and $Q_s^2=10$~GeV${}^2$, and the
momentum $k_\perp$ of the produced particle is set to $10$~GeV.
Lines: MV model. Dots: asymptotic model.}
\end{figure}

In Figure \ref{fig:Nbar-mv-as-qs}, we first compare the average number
of recoils for the two models (thin lines: MV model -- dots:
asymptotic model). The value of the ``trigger momentum'' $k_\perp$ is
held fixed at a value of $10$~GeV, and the saturation momentum squared
is varied in the range $1-10$~GeV${}^2$.  One sees that the number of
semi-hard and soft recoils is quite smaller in the asymptotic model
than in the MV model. At the largest of the considered $Q_s$, the
number of soft recoils is ten times smaller in the asymptotic model
than in the MV model. We interpret this as an effect of shadowing,
which ``hides'' the scattering centers from the passing gluon. A
similar observation was made in \cite{BlaizGV1}, where it was seen
that the multiple scatterings that lead to the Cronin effect are
almost inexistent in this asymptotic model. Also, an effect of
shadowing is that the dependence on $Q_s$ is much weaker in the
asymptotic model: piling up more and more color charges in the nucleus
does not lead to many more scatterings if the gluon cannot see them
because of shadowing.  This weaker dependence on $Q_s$ is also seen at
large $k_\perp^{\rm min}$, where one can hardly see any change even at
$Q_s^2=10$~GeV${}^2$.

Another feature of the asymptotic model is that the ``plateau'' at
$N=1$ for $Q_s \ll k_\perp^{\rm min}\lesssim k_\perp$ is no longer
really flat. Instead of a wide plateau between $Q_s$ and $k_\perp$,
one has instead a slow but steady rise of the multiplicity as
$k_\perp^{\rm min}$ decreases. For this reason, we expect the two-component 
structure of the momentum distribution of the recoils to be
less pronounced in the asymptotic model than in the MV model. This is
what we check by taking a derivative with respect to $k_\perp^{\rm
min}$, as illustrated in Figure \ref{fig:dNbar-mv-as-qs}. 
In these plots, one can see that the dip between the low-momentum
component and the peak around $k_\perp^{\rm min}=k_\perp$ is not as
deep as in the MV model. This means that one should expect the
distribution of momenta in the ``away-side jet'' to be more extended
towards softer momenta, as one probes the nucleus at smaller and
smaller values of $x$.

\section{Measuring the impact parameter\\ from the multiplicity?}
\label{sec:impact}
Based on the above study, one can address a related
question:\footnote{This question is reminiscent of the attempts to
measure the impact parameter in collisions on nuclei by counting the
so-called ``gray tracks''. Usually, in the relatively low-energy
collisions where this has been used, the picture is that the passing
projectile would kick {\sl nucleons} out of the nucleus and that by
counting these nucleons one could estimate the impact parameter. The
general idea of our study is the same, except that the action takes
place at the partonic level.} is there a correlation between the
measured multiplicity in a pA collision (event by event) and the
impact parameter of the collision?  and with what accuracy could one
determine the impact parameter based on this correlation?

In this theoretical study, the question one can answer is the
following: {\sl if the measured multiplicity in an event is $n$ (in
addition to the hard jet of momentum $k_\perp$), what is the
probability distribution of the various impact parameters?} In order
to answer this question, we will make three assumptions:
\begin{itemize}
\item[(i)] When two bunches of nuclei and protons collide in an
accelerator, all the impact parameters $\b$ are equally probable.
\item[(ii)] The only recorded events are those where $\big|\b\big|\le
  R$, where $R$ is the radius of the nucleus (we neglect the radius of
  the proton). Assuming here for simplicity that the trigger
  efficiency is the same for all $\b$'s, the probability of a given
  impact parameter (in the absence of any other information about the
  collision) is a priori equal to $P(\b)=1/\pi R^2\; .$
\item[(iii)] The density parameter $\mu_0^2$ at a given impact
parameter $b$ is proportional to the thickness of the nucleus at this
impact parameter, i.e. to $\sqrt{R^2-b^2}\; .$
\end{itemize}

Let us introduce the probability ${\cal P}(n,\b)$ of having
simultaneously the impact parameter $\b$ and the multiplicity $n$ (it
is implicit in all this section that we mean the multiplicity above a
certain threshold $k_\perp^{\rm min}$ when the passing gluon has
acquired the momentum $k_\perp$ -- these variables will not be written
anymore in order to avoid encumbering the notations).  ${\cal
P}(n,\b)$ must be normalized so that one has
\begin{equation}
\sum_{n=0}^\infty\int d^2\b\; {\cal P}(n,\b)=1\; .
\end{equation}
The probabilities $P_n$ defined earlier in this paper can be obtained
from this more general object by
\begin{equation}
P_n
=
\frac{{\cal P}(n,\b)}{\sum_n \;{\cal P}(n,\b)}\; .
\end{equation}
The denominator is necessary so that the $P_n$'s add up to
unity. Obviously, this denominator is a function that depends only on
$\b$, whose integral over $\b$ is unity. It is nothing but the
probability of having a collision with impact parameter $\b$, when the
incoming gluon has been scattered off the nucleus with a momentum
$\k_\perp$. It is easy to convince oneself that this probability is
given by:
\begin{equation}
\sum_n \;{\cal P}(n,\b)=\frac{C(\k_\perp)}{\int d^2\b\; C(\k_\perp)}\; ,
\end{equation}
where the $\b$ dependence of $C(\k_\perp)$ comes implicitly via the
parameter $\mu_0^2$. If there were no trigger bias, this quantity
would simply be uniform and equal to $(\pi R^2)^{-1}$. However,
because it is slightly more likely to have a large $\k_\perp$ in
central collisions than in peripheral ones, the mere fact of selecting
a specific $\k_\perp$ in the final state introduces a certain bias in
the distribution of impact parameters\footnote{One can check
numerically that this bias is significant only for very peripheral
collisions.}. Therefore, one has
\begin{equation}
{\cal P}(n,\b)=\frac{P_n\, C(\k_\perp)}{\int d^2\b\; C(\k_\perp)}\; ,
\end{equation}
and we see that no new calculation is necessary. It will be sufficient to
calculate $P_n$ at fixed $n$ as a function of $\b$ (the $b$ dependence
comes in via $\mu_0^2 \sim \sqrt{R^2-b^2}$).

\begin{figure}[htbp]
\begin{center}
\resizebox*{9cm}{!}{\rotatebox{-90}{\includegraphics{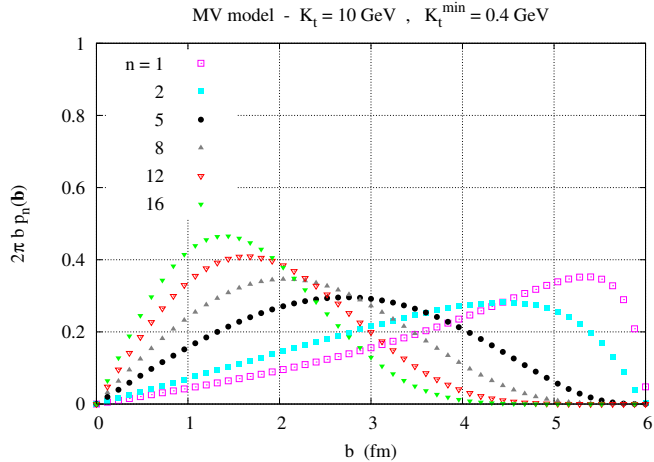}}}
\end{center}
\caption{\label{fig:b-dist}Probability distribution of the impact
parameter for various final multiplicities, in the MV model. The
trigger momentum is set to $k_\perp=10$~GeV, the threshold momentum to
$k_\perp^{\rm min}=0.4$~GeV.}
\end{figure}

From this object ${\cal P}(n,\b)$, it is easy to obtain the normalized
distribution of impact parameters conditional to having an event with
the multiplicity $n$, which is the solution to the question we asked,
\begin{equation}
p_n(\b)=\frac{{\cal P}(n,\b)}{\int d^2\b \; {\cal P}(n,\b)}\; .
\end{equation}
We have evaluated this quantity numerically in the MV model. The only
extra parameters that need to be set are the coefficient of
proportionality between the density $\mu_0^2$ and the size
$\sqrt{R^2-b^2}$ -- we set it so that the saturation scale at the
center of the nucleus ($b=0$) is $2$~GeV${}^2$ -- and the nuclear
radius, taken to be $R=6$~fm. The results are displayed in Figure
\ref{fig:b-dist}, for events where the gluon acquires the momentum
$k_\perp=10$~GeV and with a threshold momentum of $k_\perp^{\rm
min}=0.4$~GeV for counting the number of recoils.

The results are fairly intuitive: events with a low multiplicity are
dominated by large impact parameters, and events with a high
multiplicity are much more central. But we also see that selecting a
given final multiplicity only gives a fairly vague idea of the impact
parameter, since the distributions of probability for $b$ at a fixed
$n$ are quite wide, with important overlaps between the curves for
different final multiplicities. And to make things even more
difficult, the two extreme values of the final multiplicity ($n=1$ and
$n=16$ in our example), which have the least overlap in $b$,
correspond to very rare events as one can judge from the figure
\ref{fig:Pn}. Therefore, it seems realistic to make two centrality
classes, reasonably well separated in impact parameter, based on the
observed number of recoils. Changing the value of the threshold
$k_\perp^{\rm min}$ may help this separation, but we have not
investigated that approach systematically here.

\section{Conclusions}
In this paper, we have calculated the distribution of the number of
scatterings in proton-nucleus collisions, in the Color Glass
Condensate framework. This has been done by calculating the generating
function for the probabilities of having a definite number of
scatterings. We observe that, when the produced gluon has a transverse
momentum which is large compared to the saturation scale, then this
momentum is provided mostly by a single scattering center in the
nucleus, leading therefore to the familiar di-jet configuration. This
hard scattering is accompanied by a larger number of semi-hard
scatterings, with transferred momenta of the order of the saturation
momentum or smaller. By comparing the McLerran-Venugopalan model with
a model that describes the regime of very small $x$, we also see that
the shadowing present in the latter tends to suppress these semi-hard
scatterings, and to blur the separation between the hard and semi-hard
scatterings. Finally, we have discussed the correlation between the
final multiplicity and the impact parameter, and shown that it is not
a very strong correlation, that can at best be used to make a gross
separation in at most 2-3 centrality bins.

As a final note, let us mention that the results discussed in this
paper are a particular case of some general results on random walks
(in two dimensions in our case) where at each step one may have a
random step size (both in magnitude and direction), according to a
certain probability law. If this probability distribution for the step
sizes is falling very quickly, then the only way that the random walk
may end far away from the origin is to add up a very large number of
small steps. On the contrary, if this probability distribution has an
extended tail at large step sizes, such that the variance is infinite
-- such random walks are known as ``L\'evy flights'' -- then the most
efficient way to go far from the origin is to make one big step,
accompanied by smaller steps. Note that the distance from the origin
reached after a large number of steps has very different distributions
in these two situations: Gaussian in the first case, as opposed to a
power-law tail in the second case. The interested reader may see
\cite{BardoBAC1}, pp.~42-59, for a pedagogical introduction to L\'evy
statistics.

In the problem of independent multiple scatterings that we have
discussed in this paper, the ``step size'' is the transverse momentum
acquired by the gluon at each scattering, which has a probability
distribution that falls like $\sigma(\l_\perp)\sim \l_\perp^{-4}$ in
the MV model (even slower if there is an anomalous dimension different
from unity). The variance of the step sizes, $\int d^2\l_\perp
\;l_\perp^2\;\sigma(\l_\perp)$, is thus infinite, and our problem
falls in the category of L\'evy flights. Many of our results can be
understood from this analogy.

\section*{Acknowledgements}
We would like to thank J-Y. Ollitrault and P. Romatschke for
discussions on this work.

\bibliographystyle{unsrt}

\end{document}